\documentclass[12pt]{article}
  
\usepackage{amsmath,amsfonts,epsfig,mathrsfs,todonotes,yfonts}
\usepackage{xcolor}
\usepackage{cite}
\usepackage{stmaryrd}
\usepackage{mdframed}
\usepackage{scalerel}
\usepackage[top=2.5cm, bottom=2.5cm, left=2.75cm, right=2.75cm]{geometry} 
\usepackage{dsfont}
\numberwithin{equation}{section}
\usepackage{hyperref}
 \hypersetup{
     colorlinks=true,
     linkcolor=black,
     filecolor=blue,
     citecolor=blue,      
     urlcolor=cyan,
     }
\usepackage[most]{tcolorbox}
\usepackage{multirow}
\usepackage{pdflscape}
\usepackage{afterpage}

\newcommand{\re}[1] {(\ref{#1})}

\newcommand{\pa}{\partial} 

\newcommand{\ber}{\begin{eqnarray}}
\newcommand{\eer}[1]{\label{#1}\end{eqnarray}}
\newcommand{\eero}{\end{eqnarray}}

\newcommand{\balg}{\begin{align}}
\newcommand{\ealg}{\end{align}}
\newcommand{\bald}{\begin{aligned}}
\newcommand{\eald}{\end{aligned}}

\newcommand{\beq}{\begin{equation}}
\newcommand{\eeq}{\end{equation}}
\newcommand{\bea}{\begin{eqnarray}}
\newcommand{\eea}{\end{eqnarray}}

\newcommand{\nn}{\nonumber}
\newcommand{\na}{\nabla}

\newcommand{\half}{{\textstyle{\frac12}}}

\def\HollowBox #1#2{{\dimen0=#1 \advance\dimen0 by -#2
       \dimen1=#1 \advance\dimen1 by #2
        \vrule height #1 depth #2 width #2
        \vrule height 0pt depth #2 width #1
        \llap{\vrule height #1 depth -\dimen0 width \dimen1} 
       \hskip -#2
       \vrule height #1 depth #2 width #2}}
\def\BOX{\HollowBox{.100in}{.010in}}

\usepackage{fancyhdr} 
\newcommand{\auth}{\large Okan Günel ${}^{a}$\footnote{email: okan@metu.edu.tr},
{\large Ulf Lindstr\"om ${}^{a,b}$\footnote{email: ulf.lindstrom@physics.uu.se, Leverhulme Visiting Professor at Imperial College}}
and {\large \"O}zg{\"u}r Sar{\i}o\u{g}lu ${}^a$\footnote{email: sarioglu@metu.edu.tr}}

\begin{document}
\begin{flushright}
{\small UUITP-01/23}\\
\vskip 1.5 cm
\end{flushright}

\begin{center}
{\Large{\bf Killing-Yano charges of asymptotically maximally symmetric black holes}}
\vspace{.75cm}

\auth
\end{center}
\vspace{.5cm}
\centerline{${}^a${\it \small Department of Physics, Faculty of Arts and Sciences,}}
\centerline{{\it \small Middle East Technical University, 06800, Ankara, Turkey}}
\vspace{.5cm}
\centerline{${}^b${\it \small Department of Physics and Astronomy, Theoretical Physics, Uppsala University}}
\centerline{{\it \small SE-751 20 Uppsala, Sweden}}
\centerline{and}
\centerline{{\it \small Theoretical Physics, Imperial College London,}}
\centerline{{\it \small Prince Consort Road, London SW7 2AZ, UK}}

\vspace{1cm}


\centerline{{\bf Abstract}}
\bigskip

\noindent
We construct an asymptotic conserved charge for a current that has been defined using 
Killing-Yano tensors. We then calculate the corresponding conserved charges of
of the Kerr and AdS-Kerr black holes, and their higher-dimensional generalizations, 
Myers-Perry and Gibbons-L\"u-Page-Pope black holes. The new charges all turn out to be 
proportional to the angular momenta of their parent black holes.
\vskip .5cm
  
\vspace{0.5cm}
\small

\renewcommand{\thefootnote}{\arabic{footnote}}
\setcounter{footnote}{0}

\pagebreak
\tableofcontents
\setcounter{page}{2}

\section{Introduction}\label{intro}
A solution to the field equations of General Relativity may or may not support Killing 
or Killing-Yano tensors. However when it does, that is generally a sign for the
existence of ``hidden symmetries'' of the particular geometry, typically helps in the 
separation of variables in the relevant Hamilton-Jacobi equations, and usually implies the
existence of conserved currents and charges. While it is not always easy to give a clear 
physical interpretation of such conserved currents, conserved charges constructed out of 
these typically have an interpretation in terms of the parameters of the particular
solution.

One such enigmatic current, which seems to carry important information, is that of 
Kastor-Trachen (KT) \cite{Kastor:2004jk}. Finding conserved charges constructed out of this 
current for various interesting geometries would certainly help in a better understanding of 
the physical relevance of this current. The present paper is a work which, we hope, provides
a positive step in this direction.

In this paper we derive asymptotic conserved charges for the Kerr \cite{Kerr:1963ud}, the 
AdS-Kerr \cite{Boyer:1966qh} and the Gibbons-L\"u-Page-Pope (GLPP) \cite{Gibbons:2004uw} 
black holes, that are higher dimensional generalizations of the AdS-Kerr metric. The charges 
for the Myers-Perry (MP) \cite{Myers:1986un} black holes easily follow from those of the 
GLPP ones by taking the cosmological constant to zero. Our derivation uses ideas from 
\cite{Abbott:1981ff} and techniques drawn from discussions of the KT-current for 
Killing-Yano tensors (KYTs) as developed in \cite{Cebeci:2006mc} and \cite{Lindstrom:2021qrk}, 
and are further extended here.

Briefly, the procedure we follow is to split the metric into a background plus a deviation,
and consider cases where the background has a KYT. We then further restrict to the cases 
when the KT-current linearized with respect to the background is conserved and it is possible
to determine the corresponding potential for the linearized current. These then make it feasible 
to construct (asymptotically) conserved charges as integrals over sub-manifolds. By ``saturating" 
the current potential with a certain set of vectors, we can always take a flux-integral 
over a $(D-2)$-dimensional subspace. This approach is used for explicitly calculating the 
charges for the GLPP and MP black holes in $4 \leq D \leq 8$ dimensions. Our systematic 
results are tabulated in sec. \ref{hdm}. 

The organization of the paper is as follows: In sec. \ref{ktcur} we introduce KYTs and the 
KT-current. Sec. \ref{lincha} contains a review of asymptotic charges in this context,
expounds the discussion given in the previous paragraph and introduces the charge definition
employed in the rest of the paper. A treatment of the Kerr metric along these lines is contained 
in sec. \ref{KandK} and of the AdS-Kerr metric in sec. \ref{AdSKandK}. Sec. \ref{hdm} then 
presents our derivation of the charges for the GLPP and MP black holes. We end by a brief 
discussion of our results in sec. \ref{disc}.

The setting of our discussion is General Relativity in $D \geq 4$ dimensions; i.e. the geometry 
is pseudo-Riemannian with metric $g$, Levi-Civita connection $\Gamma$, curvature tensor $R$, etc. 
We also implicitly assume that the geometry supports KYTs, as described in sec. \ref{ktcur} below,
and indices of KYTs are raised and lowered with the metric $g$ that supports them in the first place.

\section{The KT-current}\label{ktcur}
Let us start by recalling some basics on KYTs. They can be thought of as
generalizations of Killing vectors to rank-$n$ antisymmetric tensor fields. So they can be
viewed as the components of $n$-forms \( f_{a_{1} \dots a_{n}} = f_{[a_{1} \dots a_{n}]} \)
satisfying 
\bea
 \nabla_{(a} f_{a_{1}) a_{2} \dots a_{n}} = 0 \,, \qquad  1 \leq n \leq D \,. \label{{Deff}} 
\eea

For a spacetime that admits a rank-$n$ KYT $f$, one can also show that the current\footnote{We 
refer to (\ref{KTcur}) as the \emph{KT-current} henceforth.} \cite{Kastor:2004jk}
\bea
j^{a_{1} \dots a_{n}} & = & N_{n} \, \delta^{a_{1} \dots a_{n} d_{1} d_{2}}_{b_{1} \dots b_{n} c_{1} c_{2}}
\, f^{b_{1} \dots b_{n}} \, R_{d_{1}d_{2}}{}^{c_{1}c_{2}}  \label{KTcur} \\
& = & - \frac{(n-1)}{4} \, R^{[a_{1} a_{2}}\,_{b c} \, f^{a_{3} \dots a_{n}] b c}
+ (-1)^{n+1} \, R_{c}\,^{[a_{1}} \, f^{a_{2} \dots a_{n}] c} - \frac{1}{2 n} \, R \, f^{a_{1} \dots a_{n}} \nn
\eea
is covariantly conserved. Here
\( \delta^{a_{1} \dots a_{m}}_{b_{1} \dots b_{m}} = 
\delta^{[a_{1}}_{b_{1}} \cdots \delta^{a_{m}]}_{b_{m}} \)
is totally antisymmetric in all up and down indices, and
\ber
N_{n} = - \frac{(n+1)(n+2)}{4n} \,.
\eer{cc}
The covariant conservation of (\ref{KTcur}), i.e., \( \nabla_{a_{1}} j^{a_{1} \dots a_{n}} = 0 \)
follows from the Bianchi identities:
\ber
&&\na_{[a}R_{bc]}{}^{de} = 0 \,, \quad 
\na_{a} R_{bcd}{}^{a} + 2 \na_{[b} R_{c]d} = 0\,, \quad 
\na_{a} R^{a}{}_{b} - \half \na_b R = 0 \,,
\eer{Bian}
and the properties of $f$.

Since a covariantly conserved antisymmetric rank-$n$ tensor field is equivalent 
to a co-closed $n$-form, one can use the extension of the Poincar\'e lemma to the exterior 
co-derivative and express the original rank-$n$ tensor field as the co-derivative of an $(n+1)$-form 
in a suitably chosen (simply-connected) open set. Thus, one should be able to write,
at least locally, 
\beq
j^{a_{1} \dots a_{n}} = \nabla_{c} \, \ell^{c a_{1} \dots a_{n}} \label{jnal}
\eeq
for some $(n+1)$-form potential \( \ell^{c a_{1} \dots a_{n}} = \ell^{[c a_{1} \dots a_{n}]} \).
The problem of determining the potential $\ell$ for the KT-current $j$ \re{KTcur} is still open.
However, this is not what we are interested in here and there is more to the story: Suppose
that one has somehow found a totally antisymmetric potential $\ell$ for the KT-current $j$ 
\re{KTcur} satisfying
\beq
\nabla_{a_{1}} \, j^{a_{1} \dots a_{n}} 
= \nabla_{a_{1}} \, \nabla_{c} \, \ell^{c a_{1} \dots a_{n}} = 0 \,. 
\label{najnanal}
\eeq
Consider a set of linearly independent arbitrary vectors \( x^{(i)} \,, (i = 1, \dots, n-1) \) in
the spacetime that admits the rank-$n$ KYT $f$ that goes into the KT-current $j$ \re{KTcur}.
Then it follows that
\beq
\nabla_{d} \, \nabla_{c} \, \big( \ell^{c d a_{1} \dots a_{n-1}} \, x^{(1)}_{a_{1}} 
\dots \, x^{(n-1)}_{a_{n-1}} \big) = 0 \,. \label{id1}
\eeq
This is easily seen if we observe that the covariant derivatives turn into a curvature tensor 
acting on a second rank antisymmetric tensor. Alternatively, to be explicit, we expand the 
derivatives on the left hand side,
\bea
&&
\left( \nabla_d \nabla_{c} \, \ell^{c d a_1 \ldots a_{n-1}} \right) x^{(1)}_{a_1} \ldots x^{(n-1)}_{a_{n-1}} 
\nn \\
&& +  \left(  \nabla_d \ell^{c d a_1 \ldots a_{n-1}} \right) \, 
\sum_{i=1}^{n-1} x^{(1)}_{a_1} \dots (\nabla_{c} x^{(i)}_{a_i}) \dots x^{(n-1)}_{a_{n-1}} 
+ (d \longleftrightarrow c) 
\nn \\
&& + \ell^{c d a_1 \ldots a_{n-1}} \, \sum_{i=1}^{n-1} \, \sum_{j \neq i}^{n-1} 
x^{(1)}_{a_1} \dots (\nabla_{c} x^{(i)}_{a_i}) \dots (\nabla_{d} x^{(j)}_{a_j}) \dots x^{(n-1)}_{a_{n-1}} 
\nn \\
&& + \ell^{c d a_1 \ldots a_{n-1}} \, \sum_{i=1}^{n-1} x^{(1)}_{a_1} \dots 
\left( \nabla_{d} \nabla_{c} x^{(i)}_{a_i} \right) \dots x^{(n-1)}_{a_{n-1}} \,. \label{acac}
\eea
The first term in \re{acac} vanishes by \re{najnanal}. The terms on the second and third lines 
are symmetric on the indices $c$ and $d$, whereas $\ell$ itself is totally antisymmetric; so they 
vanish. The last term can be rewritten in terms of the Riemann tensor by using the antisymmetry 
of $\ell$ again:
\bea
\ell^{c d a_1 \ldots a_{n-1}} \sum_{i=1}^{n-1} x^{(1)}_{a_1} \dots 
\left( \nabla_{d} \nabla_{c} x^{(i)}_{a_i} \right) \dots x^{(n-1)}_{a_{n-1}} =
2 \ell^{c d a_1 \ldots a_{n-1}} \sum_{i=1}^{n-1} x^{(1)}_{a_1} \dots 
\left( R_{d c a_{i}}{}^{b} \, x^{(i)}_{b} \right) \dots x^{(n-1)}_{a_{n-1}},
\eea
which vanishes by the Bianchi identity \( R_{[d c a_{i}]}{}^{b} = 0 \). Hence \re{id1} does hold
whenever a spacetime admits a rank-$n$ KYT $f$ and there is a proper set of $(n-1)$ vectors 
$x^{(i)}$. 

In the next section we will discuss how these can be used for defining conserved charges. 

\section{Linearized currents and asymptotic charges}\label{lincha}
The existence of asymptotic charges based on the KT-current \re{KTcur} was shown 
in \cite{Kastor:2004jk,Cebeci:2006mc} for asymptotically flat and asymptotically AdS 
geometries. The method is a generalization of employing asymptotic 
Killing vectors \cite{Abbott:1981ff} to define the corresponding conserved charges.
Let us quickly recapitulate this construction for convenience. The starting point is a 
$D$-dimensional spacetime with a metric $g_{ab}$ whose asymptotic Killing-Yano charge(s) 
are to be computed. The metric $g_{ab}$ does not necessarily have to admit exact KYTs, but 
the assumption is that it can be split into a background $\bar{g}_{ab}$ plus a deviation as
\beq
g_{ab} \equiv \bar{g}_{ab} + h_{ab} \quad \mbox{so that} \quad 
g^{ab} = \bar{g}^{ab} - h^{ab} + {\cal O}(h^2) \,, \label{metspl}
\eeq 
where $h^{ab}=\bar{g}^{ac}h_{cd}~\!\bar{g}^{db}$. It is also assumed that
$h_{ab}$ vanishes sufficiently fast at the relevant (spatial) boundary, and that $\bar{g}_{ab}$
admits a completely antisymmetric rank-$n$ KYT $\bar{f}_{a_{1} \dots a_{n}}$.

With the understanding that all indices are raised and lowered with the generic background 
metric $\bar{g}_{ab}$ from now on, e.g. \( h \equiv \bar g^{ab} h_{bc} \) and 
\( \bar{\BOX} \equiv {\bar{\nabla}}^{c} \, {\bar{\nabla}}_{c} \), one finds the following linearized 
curvature, Ricci tensor and curvature scalar (to ${\cal O}(h)$)
\ber
( R_{ab}{}^{cd} )_{L} & = & \bar{R}_{abe}{}^{[c} h^{d]e} + 2 \, \bar{\na}_{[a} \bar{\na}^{[d} h_{b]}{}^{c]} 
\,, \label{RiemL} \\[1mm]
( R^{a}{}_{b} )_{L} & = & \frac{1}{2} \left( \bar{\na}^{c} \bar{\na}^{a} h_{bc} + 
\bar{\na}_{c} \bar{\na}_{b} h^{ac} - \bar{\na}^{a} \bar{\na}_{b} h - \bar{\BOX} h^{a}{}_{b} \right)
- h^{ac} \bar{R}_{bc} \,,
\\[1mm]
R_{L} & = & {\bar{\nabla}}_{a} {\bar{\nabla}}_{b} h^{ab} - \bar{\BOX} h - h^{ab} \bar{R}_{ab} \,.
\eer{Lin}
For maximally symmetric backgrounds that we will be concerned with, the 
linearized Bianchi identities are identical to \re{Bian} with all curvature terms replaced by their 
linearized counterparts. This, in turn, guarantees the background conservation of the linearized 
current (see \cite{Lindstrom:2021qrk} for technical details), i.e. 
\( \bar{\nabla}_{a_{1}} (j^{a_{1} \dots a_{n}})_{L} = 0 \). 
Since the current is antisymmetric, the covariant conservation gives rise to an ordinary 
conservation law via
\beq
\bar{\nabla}_{a_{1}} (j^{a_{1} \dots a_{n}})_{L} = \frac{1}{\sqrt{|\bar{g}|}} \,
\partial_{a} \big( \sqrt{|\bar{g}|} \, (j^{a_{1} \dots a_{n}})_{L} \big) = 0 \,. \label{conslaw}
\eeq
This is used in constructing conserved charges for linearized currents in 
\cite{Kastor:2004jk,Cebeci:2006mc}, as flux integrals over a suitably chosen 
$(D-1-n)$-dimensional subspace by further invoking Stokes' theorem: The crucial 
step is the determination of a potential $\bar{\ell}$ for the current $j_{L}$, as further 
elucidated below.

The asymptotic charges for the KT-current were given in \cite{Kastor:2004jk} for an arbitrary 
rank-$n$ KYT in an asymptotically flat background and in \cite{Cebeci:2006mc} for an arbitrary 
rank-$n$ KYT in an asymptotically AdS background. As mentioned, their existence rests on 
the linearized KT-current  being expressible as the covariant divergence of an $(n+1)$-form 
potential. The construction of such a potential is nontrivial, a condition that the background 
has to satisfy for this procedure to work was derived in \cite{Lindstrom:2021qrk}. 

The relevant linearized part of \re{KTcur} is
\ber
(j^{a_{1} \dots a_{n}})_{L} = N_{n} \, 
\delta^{a_{1} \dots a_{n} d_{1} d_{2}}_{b_{1} \dots b_{n} c_{1} c_{2}} \, 
\bar{f}^{b_{1} \dots b_{n}} \, (R_{d_{1}d_{2}}{}^{c_{1}c_{2}})_{L} \,.
\eer{KTComL}
In terms of the linearized Riemann tensor in \re{RiemL}, the current may be written as
\ber
(j^{a_{1} \dots a_{n}})_{L} = N_{n} \, 
\delta^{a_{1} \dots a_{n} d_{1} d_{2}}_{b_{1} \dots b_{n} c_{1} c_{2}} \, \bar{f}^{b_{1} \dots b_{n}} \,
\big( \bar{R}_{d_{1} d_{2} e}{}^{[c_{1}} \, h^{c_{2}] e} 
+ 2 \, \bar{\na}_{d_{1}} \bar{\na}^{c_{2}} h_{d_{2}}{}^{c_{1}} \big) \,.
\eer{dd}
For a flat background, this may be written as \cite{Kastor:2004jk} 
\ber
(j^{a_{1} \dots a_{n}})_{L}  = \bar{\nabla}_{c} \, \bar{\ell}^{c a_{1} \dots a_{n}} = 
\frac{1}{\sqrt{|\bar{g}|}} \pa_{c} \Big( \sqrt{|\bar{g}|} \, \bar{\ell}^{c a_{1} \dots a_{n}} \Big)
\eer{ee}
where the $(n+1)$-form \( \bar{\ell}^{e a_{1} \dots a_{n}} = \bar{\ell}^{[e a_{1} \dots a_{n}]} \) is
\ber
\bar{\ell}^{e a_{1} \dots a_{n}} = 2 N_{n} \, 
\delta^{a_{1} \dots a_{n} e d_{2}}_{b_{1} \dots b_{n} c_{1} c_{2}} \,
\bar{f}^{b_{1} \dots b_{n}} \, \bar{\nabla}^{c_2} \, h_{d_{2}}{}^{c_{1}}
- \frac{1}{2n} \Big( h \, \bar{\nabla}^{e} \, \bar{f}^{a_{1} \dots a_{n}}
- (n+1) \, h^{d_{2} [e} \, \bar{\nabla}_{d_{2}} \, \bar{f}^{a_{1} \dots a_{n}]} \Big) .
\eer{ell}
Similar manipulations as in \cite{Kastor:2004jk} give the following result for the general 
case\footnote{Note that there are no additional curvature terms generated in the process.}
\cite{Lindstrom:2021qrk} 
\ber
(j^{a_{1} \dots a_{n}})_{L} = \bar{\nabla}_{e} \, \bar{\ell}^{e a_{1} \dots a_{n}}
+ N_{n} \Big( \bar{f}^{[a_{1} \dots a_{n}} \, \bar{R}_{c_{1} c_{2} e}{}^{c_{1}} \, h^{c_{2}]e}
+ 2 \, h_{c_{2}}{}^{[c_{1}} \, \bar{\nabla}^{c_{2}} \, 
\bar{\nabla}_{c_{1}}  \, \bar{f}^{a_{1} \dots a_{n}]} \Big)
\eer{Rdiff}
with $\bar{\ell}$ as in \re{ell}.
It was shown in \cite{Lindstrom:2021qrk} that the vanishing of the terms in parenthesis 
\re{Rdiff} can be expressed in terms of the background curvature as\footnote{When $n=1$,
\re{ADcnd} simply reads \( h \bar{R}^{ab} \bar{f}_{b} - h^{bc} \bar{R}_{bc} \bar{f}^{a} = 0 \).}
\ber
 \bar{f}^{[a_{1} \dots a_{n}} \, \bar{R}_{c_{1} c_{2} e}{}^{c_{1}} \, h^{c_{2}]e}
  + 2 (-1)^{n} h_{c_{2}}{}^{[c_{2}} \bar{R}_{e}{}^{c_{1}}{}_{c_{1}}{}^{a_{1}} \, 
 \bar{f}^{a_{2} \dots a_{n}] e} = 0 \,.
\eer{ADcnd}
It is clearly fulfilled for the flat case which leads to the results in \cite{Kastor:2004jk}.
For a maximally symmetric background 
\[ \bar{R}_{abcd} = \lambda \, (\bar{g}_{ac} \, \bar{g}_{bd} - \bar{g}_{ad} \, \bar{g}_{bc}) \,, \quad
\bar{R}_{ab} =  (D-1) \lambda \, \bar{g}_{ab} \,, \quad
\bar{R} = D (D-1) \lambda \,, \]
\re{ADcnd} is also fulfilled and leads to the results in \cite{Cebeci:2006mc}. This agrees 
with the known cases where the linearized Bianchi identities ensure conservation 
of the KT-current. 

To give a concrete example, let us explicitly write the potential for the rank-$2$ case, 
i.e. \( (j^{ab})_{L}  = \bar{\nabla}_{c} \, \bar{\ell}^{cab} \)
for a maximally symmetric background \cite{Kastor:2004jk,Cebeci:2006mc}:
\ber
\bar{\ell}^{abc} = -\frac{3}{2}\, \bar{f}^{d[a} \, {\bar{\nabla}}^{b} \, h^{c]}\,_{d} 
+ \frac{3}{4} \, \bar{f}^{[ab} \, {\bar{\nabla}}^{c]} \, h 
+ \frac{3}{4} \, h^{d[c} \, {\bar{\nabla}}_{d} \, \bar{f}^{ab]} 
- \frac{3}{4} \, \bar{f}^{[ab} \, {\bar{\nabla}}_{d} \, h^{c]d}
- \frac{1}{4} \, h \, {\bar{\nabla}}^{[a} \, \bar{f}^{bc]} \,.
\eer{yanc}
Furthermore, the GLPP \cite{Gibbons:2004uw} or the MP \cite{Myers:1986un} black 
holes we will consider in this work clearly have maximally symmetric backgrounds, and
all can be cast into the form \cite{Kastor:2004jk}
\ber
\bar{g}_{ab} = (n^c n_c) n_a n_b + r_a r_b + q_{ab} \,,
\eer{back}
where $q_{ab}$ is the metric on the $(D-2)$-dimensional space $\Sigma$, $n^a$ and $r^a$ are 
mutually orthogonal unit vectors to $\Sigma$, with $n^a$ a non-null vector, which we will choose 
to be timelike so that \( n^c n_c = -1\). 
The conserved charge can be written for the $n=2$ case using the $3$-form potential \re{yanc} as
\ber
Q = \int_{\Sigma} d^{D-2} x \, n_{[a} \, r_{b]} \, \big( \sqrt{|q|} \, (j^{ab})_{L} \big) = 
\int_{\Sigma} d^{D-2} x \, n_{[a} \, r_{b]} \, \pa_{c} \big( \sqrt{|\bar{g}|} \, \bar{\ell}^{cab} \big).
\eer{bb} 
Near the spatial boundary, one can further write \( q_{ab} = y_a y_b + \gamma_{ab} \), where 
at spatial infinity $y^a$ is the unit normal to the $(D-3)$-dimensional boundary $\pa \Sigma$ 
and $\gamma_{ab}$ is the metric on $\pa \Sigma$. Finally, Stokes' theorem lets one write
\ber
Q = \int_{\pa \Sigma} d^{D-3} x \, n_{[a} \, r_{b} \, y_{c]} \, \big( \sqrt{|\gamma|} \, 
\bar{\ell}^{abc} \big)
\eer{QQ}
at the spatial boundary. 

However, one can define a conserved charge in an alternative way. In view of 
\re{conslaw} and \re{ee}, it is easy to see that the potential $\bar{\ell}$ satisfies
\beq
\label{poteq}
\bar{\nabla}_d \bar{\nabla}_{c} \, \bar{\ell}^{c d a_{1} \dots a_{n-1}} = 0 \,.
\eeq
This means that the linearization process can also be applied to \re{id1} resulting in its
linearized analog
\beq
\bar{\nabla}_d \bar{\nabla}_{c} \, 
\left( \bar{\ell}^{c d a_{1} \dots a_{n-1}} x^{(1)}_{a_1} \ldots x^{(n-1)}_{a_{n-1}} \right) = 0 \,, 
\label{leq1}  
\eeq
for a suitably chosen set of vectors \( x^{(i)} \,, (i = 1, \dots, n-1) \) on the background geometry.
The term inside the parentheses in \re{leq1} can be thought of as a 2-form ``potential" 
$\bar{L}^{cd}$, for a 1-form ``conserved current" $\bar{J}^{d}$, with 
\( \bar{J}^{d} = \bar{\nabla}_{c} \, \bar{L}^{cd} \), which can be
employed in a consistent definition for a ``conserved charge". What about a simple but 
reasonable choice for the background vectors $x^{(i)}$ though? Suppose that the background 
space admits the foliation\footnote{This is indeed the case for the maximally symmetric 
backgrounds of the MP \cite{Myers:1986un} and the GLPP \cite{Gibbons:2004uw} 
black holes we will consider.}
\beq
\bar{g}_{a b} = - n_a n_b + r_a r_b + q_{ab} \quad \mbox{with} \quad 
q_{ab} = \sum_{i=1}^{n-1} x^{(i)}_a x^{(i)}_b + \gamma_{a b} \,, 
\label{fol}
\eeq
where for consistency it must be that \( D - 2 = (n-1) + {\rm dim} \, \gamma \), so that 
\( n+1 \leq D \). Here it is implicitly assumed that the induced metric $\gamma$ on the
$(D-1-n)$-dimensional subspace is non-degenerate. Now let $x^{(i)}$ be mutually
orthogonal spacelike normal vectors of this subspace, i.e. \( x^{(i)}{}_{a} x^{(j)}{}^{a} = 0 \)
when \( i \neq j \). In fact, we also demand that they are hypersurface orthogonal, i.e.
\( x^{(i)}{}_{[a} {\bar{\nabla}}_{b} x^{(i)}{}_{c]} = 0 \) for all \( i = 1, \dots, n-1 \). With these,
one can now integrate over the $(D-2)$-dimensional spatial boundary\footnote{instead of 
integrating over the $(D-1-n)$-dimensional subspace as in \re{QQ}.} to get a 
conserved charge, and write
\beq
{\cal Q} = \int_{\Sigma} d^{D-2}x \, n_{[a} \, r_{b} \,
x^{(1)}_{c_{1}} \dots x^{(n-1)}_{c_{n-1}]} \, \bar{\ell}^{a b c_1 \dots c_{n-1}} \sqrt{|q|} \,.
\label{theQ}
\eeq

In what follows we will calculate this charge ${\cal Q}$ for Kerr, AdS-Kerr, and the 
GLPP \cite{Gibbons:2004uw} black holes, which are higher dimensional generalizations of the 
AdS-Kerr metric. The charges for the MP \cite{Myers:1986un} black holes will follow from those 
of the GLPP ones by taking the cosmological constant to zero, of course.
 
\section{The Kerr metric}\label{KandK}
Let us examine all of these ideas first on the celebrated Kerr metric \cite{Kerr:1963ud} cast in 
Boyer-Lindquist coordinates \cite{Boyer:1966qh}:
\bea
ds^{2} &=&  - \left( 1- \frac{2 M r}{\Sigma} \right) dt^{2} 
-\frac{4 a M r \sin^2{\theta}}{\Sigma} \, dt \, d\phi + \frac{\Sigma}{\Delta} \, dr^{2} + \Sigma \, d\theta^2 
\nn \\
& & + \left( r^2 + a^2 + \frac{2 a^2 M r \sin^2{\theta}}{\Sigma} \right) \sin^2{\theta} \, d\phi^{2} \,,
\label{Kerr} 
\eea
where
\beq
\Sigma := r^{2} + a^{2} \cos^{2}\theta \qquad \mbox{and} \qquad
\Delta := r^{2} + a^{2} - 2 M r \,. \label{SigDel}
\eeq
The rank-2 KYT of the Kerr metric is \cite{Floyd,Penrose:1973um}
\beq
f = a \cos{\theta} \, dr \wedge \big( dt - a \sin^{2}{\theta} \, d\phi \big)
+ r \sin{\theta} \, d\theta \wedge \big( (r^{2} + a^{2}) d\phi - a dt \big) \,. 
\label{KY2Kerr}
\eeq

The background is found by setting \( M=0, a=0 \) in \re{Kerr}:
\bea
\bar{g}_{ab} & = & {\rm diag} (-1, 1, r^2, r^2 \sin^{2}\theta) \,, \;\;
n^a = (-\pa_{t})^a \,, r^a = (\pa_{r})^a \; \mbox{with} \; n^a n_a = -1 \; \mbox{and} \; r^a r_a = 1 \,, 
\nn \\
q_{ab} & = & {\rm diag} (0, 0, r^2, r^2 \sin^{2}{\theta}) \,. \label{KNback}
\eea
Clearly $q_{ab} $ is the metric on a sphere $S^2$ of radius $r$. The spatial boundary 
is defined by \( r \to \infty \). From \re{KY2Kerr}, the background KYT reads 
\( \bar{f} = r^{3} \sin{\theta} \, d\theta \wedge d\phi \), \( \sqrt{|q|} = r^2 \sin{\theta} \) 
and after some calculation
\ber
\bar{\ell}^{tr\theta} = 
\frac{a M \sin{\theta} \left(a^2 \cos^2{\theta} +3 r^2\right)}{2 r \Sigma^2} \,.
\eer{lKN}
With the choice \( x^{a} = (\pa_{\theta}/r)^a \) from \re{fol}, it follows that
\bea
{\cal Q} & = & \lim_{r \to \infty} 
\int_{S^2} d^2 x \, n_{[a} \, r_{b} \, x_{c]} \bar{\ell}^{a b c} \sqrt{|q|} 
= \lim_{r \to \infty} \int_{0}^{\pi} d\theta \, \int_{0}^{2\pi} d\phi \, r^3 \sin{\theta} \, \bar{\ell}^{tr\theta} \nn \\
& = & \lim_{r \to \infty} 
\left( \tfrac{\pi^{2} a M r \left(2 a^6+16 r^5 \left(r-\sqrt{a^2+r^2}\right)-4 a^2 r^3 \left(7 \sqrt{a^2+r^2}-9 r\right)+3 a^4 r \left(7 r-3 \sqrt{a^2+r^2}\right)\right)}{\sqrt{a^2+r^2} \left(a^3+2 a r \left(r-\sqrt{a^2+r^2}\right)\right)^2} \right) \nn \\
& = & \frac{3 \pi^2}{2} a M \,. \label{KerrQ}
\eea
${\cal Q}$ \re{KerrQ} is proportional to $a M$, i.e. the angular momentum of the Kerr black hole.

\section{The AdS-Kerr metric}\label{AdSKandK}
In Boyer-Lindquist coordinates the AdS-Kerr metric \cite{Carter:1968ks} is
\bea
\label{eqn:KerrAdS}
ds^{2} & = & - \frac{\Delta_\theta}{\Xi} (1- \lambda r^2) \, dt^2 + \frac{\Sigma}{\Delta} \, dr^{2} + \frac{\Sigma}{\Delta_\theta} \, d\theta^2 \nn \\
& & + \frac{(r^2+a^2)}{\Xi} \sin^2{\theta} \, d\phi^2 
+ \frac{2 M r}{\Sigma} \left( \frac{\Delta_\theta}{\Xi} \, dt - \frac{a \sin^2{\theta}}{\Xi} \, d\phi \right)^2 \,,
\eea
where
\bea
&& \Delta := (r^2+a^2)(1- \lambda r^2) - 2 M r \,,  \qquad  \Xi := 1+ \lambda a^2 \,, \nn \\
&& \Delta_\theta := 1+ \lambda a^2 \cos^2{\theta} \,, \qquad \Sigma := r^2 + a^2 \cos^2{\theta} \,.
\eea
The rank-2 KYT of the AdS-Kerr metric \re{eqn:KerrAdS} is
\beq
f = \frac{a \cos{\theta}}{\Xi} \, dr \wedge \big( \Delta_\theta \, dt - a \sin^{2}{\theta} \, d\phi \big)
+ \frac{r \sin{\theta}}{\Xi} \, d\theta \wedge \big( (r^{2} + a^{2}) d\phi - a (1- \lambda r^2) dt \big) \,. 
\label{KY2AdSKerr}
\eeq
The AdS background
\beq
\bar{g}_{ab} = {\rm diag} \left(-(1- \lambda r^2), \frac{1}{1- \lambda r^2}, r^2, r^2 \sin^{2}{\theta} \right)
\eeq
is obtained by setting \( M=0,\, a=0 \) in \re{eqn:KerrAdS}. This background can be foliated as
in \re{fol}, where the unit normal vectors $n^a$, $r^a$ and $x^a$ are explicitly
\beq 
n^{a} = \left( - \frac{1}{\sqrt{1-\lambda r^2}} \partial_t \right)^a \,, \quad
r^{a} = \left( \sqrt{1-\lambda r^2} \, \partial_r \right)^a \,, \quad
x^{a} = \left( \frac{\partial_\theta}{r} \right)^a \,,
\eeq
with $\sqrt{|q|} = r^2 \sin{\theta}$. The background KYT reads 
\( \bar{f} = r^3 \sin{\theta} d\theta \wedge d\phi \) as for the Kerr case 
and leads to the potential with a nontrivial component
\beq
\bar{\ell}^{tr\theta} = 
\frac{a M \sin{\theta} \Delta_\theta \left(a^2 \cos^2{\theta} + 3 r^2 \right)}{2 r \Xi^2 \Sigma^2} \,.
\eeq
The conserved charge can be obtained by an integration on the $(D-2)$-dimensional boundary 
$S^2$ at the spatial infinity $r \rightarrow \infty$:
\beq
{\cal Q} = \lim_{r \to \infty} 
\int_{S^2} d^2 x \, n_{[a} \, r_{b} \, x_{c]} \bar{\ell}^{a b c} \sqrt{|q|} \\
= \frac{3 \pi^2 \left(3+\Xi \right)}{8 \Xi^2} a M \,, 
\label{AdSKerrQ}
\eeq
which correctly reduces to \re{KerrQ} when the cosmological constant $\lambda \to 0$. Note also
that ${\cal Q}$ \re{AdSKerrQ} is again proportional to $a M$, the angular momentum of the 
AdS-Kerr black hole.

\section{Higher-dimensional metrics}\label{hdm}
The charge ${\cal Q}$ \re{theQ} can be computed for the GLPP metrics \cite{Gibbons:2004uw} 
in higher dimensions. To that end, we first recall their form (as they are cast in Appendix E 
of \cite{Gibbons:2004uw} in Boyer-Lindquist coordinates):
\bea
\label{eqn:GLLPmet}
ds^2 & = & -W (1-\lambda r^2 ) \, dt^2 
+ \frac{2M}{VF} \left( W \, dt - \sum_{i=1}^{N} \frac{a_i \mu_i^2}{\Xi_i} \, d\phi_i \right)^2 
+ \sum_{i=1}^{N} \frac{r^2 + a_i^2}{\Xi_i} \mu_i^2 \, d\phi_i^2  \nn \\
& & + \frac{VF}{V-2M} \, dr^2 + \sum_{i=1}^{N+\epsilon} \frac{r^2 + a_i^2 }{\Xi_i} \, d\mu_i^2 
+ \frac{\lambda}{W (1-\lambda r^2)} 
\left( \sum_{i=1}^{N+\epsilon} \frac{r^2 + a_i^2}{\Xi_i} \mu_i \, d\mu_i \right)^2 \,,
\eea
where
\bea
&& W := \sum_{i=1}^{N+\epsilon} \frac{\mu_i^2}{\Xi_i} \,, \qquad 
V := r^{\epsilon-2} (1-\lambda r^2) \prod_{i=1}^N (r^2 + a_i^2) \,, \nn \\
&& F := \frac{r^2}{1-\lambda r^2} \sum_{i=1}^{N+\epsilon} \frac{\mu_i^2}{r^2 + a_i^2} \,, 
\qquad \Xi_i := 1+\lambda a_i^2  \,.
\eea
The ``evenness" integer\footnote{\( \epsilon = 1 \) 
for even $D$ and \( \epsilon = 0 \) for odd $D$.} $\epsilon$ and the number of azimuthal 
angular coordinates $N$ is defined as \cite{Gibbons:2004uw} 
\[ \epsilon := (D-1) \; \mbox{mod} \; 2 \,, \qquad  N := [(D-1)/2] \,, \]
so that the number of latitudinal coordinates $\mu_i$ is 
$N+\epsilon$, and there are $N$ azimuthal angular coordinates $\phi_j$.
Thus \( D = 2N+\epsilon+1 \), where the finger counting goes as follows: There are the time 
coordinate $t$, the radial coordinate $r$, $N+\epsilon-1$ independent latitudinal
coordinates\footnote{There is a constraint on latitudinal coordinates: 
\( \sum_{i=1}^{N+\epsilon} \mu_i^2 = 1\).} $\mu_i$ , and  
$N$ azimuthal angular coordinates $\phi_j$. The latitudinal coordinates $\mu_i$ range
over $[0,1]$ except for the last one in \emph{even} dimensions, $\mu_{N+1}$, which 
ranges over $[-1,1]$. As usual, the azimuthal angular coordinates $\phi_j$ are periodic with 
period $2\pi$. The GLPP metric \re{eqn:GLLPmet} satisfies the cosmological Einstein equations
\cite{Gibbons:2004uw} 
\[ R_{a b} = (D-1) \lambda g_{a b} \,, \]
and reduces to the MP metric \cite{Myers:1986un} when $\lambda \rightarrow 0$.
Moreover, the rank-2 closed conformal KYT of the GLPP metric (\ref{eqn:GLLPmet}) 
is\cite{Kubiznak:2008qp}
\footnote{Initially we thought our work was the first where this was reported. However we 
became aware of the thesis \cite{Kubiznak:2008qp} (see formula (B.15) therein) after posting the 
first version of our paper. In passing, it should be mentioned that the associated KYTs 
for the MP black holes were first reported in \cite{Frolov:2006dqt}, which paved the way 
for applications of KYTs to higher dimensional stationary black holes.}
\beq
k = \sum_{i=1}^{N} a_i \, \mu_i \, d\mu_i \wedge 
\left[  a_i \, dt - \frac{(r^2 + a_i^2)}{\Xi_i} \left( d\phi_i - \lambda a_i \, dt \right) \right] 
+ r dr \wedge \left[ dt - \sum_{i=1}^{N} \frac{a_i \mu_i^2}{\Xi_i} 
\left( d\phi_i + \lambda a_i \, dt \right) \right] \,.
\label{CCKYT}
\eeq

The background can be found by setting $M=0$ and $a_i=0$ in (\ref{eqn:GLLPmet}):
\bea
d\bar{s}^2 = - (1-\lambda r^2 ) \, dt^2 + \frac{dr^2}{1-\lambda r^2} 
+ r^2 \left( \sum_{i=1}^{N+\epsilon} d\mu_i^2 + \sum_{i=1}^{N} \mu_i^2 \, d\phi_i^2 \right) \,.
\label{backG}
\eea
However, this is a bit misleading since there is a constraint on the latitudinal coordinates $\mu_i$
whose elimination resurrects the $\bar{g}_{\mu_{i} \mu_{j}}$ components. This 
considerably complicates the charge ${\cal Q}$ integration \re{theQ}, since with a non-diagonal
$q_{ab}$ \re{fol}, it is difficult to identify a proper set of vectors $x^{(i)}$ that heeds 
the requirements mentioned in the penultimate paragraph of sec. \ref{lincha}. The remedy is 
a transformation from $\mu_i$ to the quasi-spheroidal coordinates $\theta_i$, 
which diagonalize the $(D-2)$-dimensional $q_{ab}$ in the foliation of the background \re{fol},
and are given  by
\beq
\mu_i(\theta) := \left( \prod_{j=1}^{N+\epsilon-i} \sin{\theta_{j}} \right) \cos{\theta_{N+\epsilon-i+1}} \,,
\label{muthe}
\eeq
where the last coordinate $\theta_{N+\epsilon}$ and $d\theta_{N+\epsilon}$ are set to 0 in order 
to write the transformation compactly. All $\theta_i$ range over $[0,\pi/2]$, except for 
$\theta_N$ in \emph{even} dimensions which ranges over $[0,\pi]$.

In the new coordinates $(t, r, \theta_i, \phi_j)$, the background metric \re{backG} is
\beq
\bar{g}_{ab} = {\rm diag} \left( -(1- \lambda r^2) \,, \frac{1}{1-\lambda r^2} \,, 
r^2 \left( \prod_{k=1}^{N+\epsilon-1-i} \sin^2{\theta_{N+\epsilon-k}} \right) \,,
r^2 \mu_j^2(\theta) \right) \,,
\label{backnG}
\eeq
where $i$ runs from 1 to $N+\epsilon-1$, $j$ runs from 1 to $N$ and we have 
kept \( \mu_j = \mu_j(\theta) \) \re{muthe} in the last entry for economy of notation.
Now, $\bar{g}_{ab}$ \re{backnG} can be put into the form (\ref{fol}) with the timelike 
and spacelike unit normal vectors
\beq
n^{a} = \left( - \frac{1}{\sqrt{1-\lambda r^2}} \partial_t \right)^a \,, \quad
r^{a} = \left( \sqrt{1-\lambda r^2} \, \partial_r \right)^a \,,
\eeq
and $q$ becomes the metric on $S^{D-2}$, the $(D-2)$-dimensional sphere, of radius $r$, 
where the spatial boundary is defined by $r \rightarrow \infty$. For this background, 
the rank-2 closed conformal KYT \re{CCKYT} simplifies to
\beq
\label{eqn:backgCCKYT}
\bar{k} = r dr \wedge dt \ .
\eeq
The rank-$(D-2)$ background KYT $\bar{f} $ that will go into the rank-$(D-1)$ potential $\bar{\ell}$ 
\re{ell} can be found by taking the Hodge dual of $\bar{k}$ (\ref{eqn:backgCCKYT}) 
\cite{Cariglia:2003kf} with respect to the background \re{backnG}
\beq
\label{eqn:backgKYT}
\bar{f} = \star \bar{k} := r \sqrt{|\bar{g}|} \, d\theta_1 \wedge \dots \wedge d\theta_{N+\epsilon-1} 
\wedge d\phi_1 \wedge \dots \wedge d\phi_N \,.
\eeq
Finding the potential $\bar{\ell}$ \re{ell} is not an easy feat now since the coordinate transformation
we have introduced earlier \re{muthe} changes the form of \re{eqn:GLLPmet} drastically, which
in turn changes the deviations $h_{ab}$ \re{metspl} that go into $\bar{\ell}$ \re{ell}. In retrospect,
a choice had to be made in the trade-off between ``being able to integrate with simpler set of 
vectors $x^{(i)}$" and ``working with more straightforwardly calculable (and perhaps less 
complicated) deviations $h_{ab}$, and hence potential $\bar{\ell}$". We have opted for the first 
and paid a price in the determination of the deviations and later the potentials. The calculations 
involved are hardly illuminating, so we prefer not to show the gory details, and briefly summarize 
the steps taken instead: First, we have cast \re{eqn:GLLPmet} in the $(t, r, \theta_i, \phi_j)$ 
coordinates for dimensions $4\leq D \leq 8$, then determined $h_{ab}$ \re{metspl} in each case 
using the background $\bar{g}_{ab}$ \re{backnG} and carefully calculated the rank-$(D-1)$ 
potential $\bar{\ell}$ \re{ell} in each case. We have found that there are $N$ independent 
components of $\bar{\ell}$ that can be used in the charge ${\cal Q}$ integration \re{theQ}: 
\beq
\bar{\ell}^{t r \theta_1 \dots \theta_{N+\epsilon - 1} \phi_1 \dots \widehat{{\phi}_{j}} \dots \phi_N} \,,
\quad ( 2N+\epsilon-1 = D-2 ) \,,
\eeq
where a wide hat on a ${\phi}_{j}$-component indicates that it is to be omitted and \(j = 1, \dots, N\). 
So we have chosen the $N+\epsilon-1$ vectors $x^{(i)}$ and the $N$ vectors $y^{(i)}$ that will 
saturate the $(D-2)$ components of the rank-$(D-1)$ potential $\bar{\ell}$ as
\beq
x^{(i)} = r \left( \prod_{k=1}^{N+\epsilon-1-i} \sin{\theta_{N+\epsilon-k}} \right) d \theta_i 
\quad \mbox{and} \quad 
y^{(j)} = r \, \mu_{j}(\theta) \, d\phi_j \,.
\eeq
Thus we have been able to calculate $N$ charges \re{theQ} in $D$ dimensions as
\beq
{\cal Q}^{(j)}_{D} = \int_{S_{D-2}} d^{D-2} x \, 
n_{[a} \, r_{b} \, x^{(1)}_{c_1} \dots x^{(N+\epsilon-1)}_{c_{N+\epsilon-1}} \, 
y^{(1)}_{c_{N+\epsilon}} \dots \widehat{y^{(j)}_{c_{N+\epsilon+i-1}}} \dots y^{(N)}_{c_{D-3}]}  \,
\bar{\ell}^{a b c_1 \dots c_{D-3}} \, \sqrt{|q|}
\eeq
in the limit $|r| \rightarrow \infty$. Our findings are tabulated in Table \ref{ResTable}. 

\afterpage{
\begin{landscape}
\begin{table}[ht!]
\centering    
\renewcommand{\arraystretch}{1.7}
\caption{Charges ${\cal Q}$ for GLLP and MP black holes in dimensions $4 \leq D \leq 8$}
\label{ResTable}
\begin{tabular}{c| c| c| c| c} 
 D &  Charge & Potential & GLLP & MP \\ [0.5ex] 
 \hline \hline
\multirow{1}{*}{4} & $Q^{(1)}_{4}$ & $\ell^{a b c_1} x^{(1)}_{c_1}$ & $\displaystyle \frac{3 a M \pi ^2 }{8 \Xi^2} (3+\Xi)$ &  $\displaystyle \frac{3 a M \pi ^2 }{2}$ \\[0.5ex]
 \hline
  \multirow{2}{*}{5} & $Q^{(1)}_{5}$ & $\ell^{a b c_1 c_2 } x^{(1)}_{c_1} y^{(2)}_{c_2}$ &  $\displaystyle  -\frac{16 a_1 M \pi ^2 }{45 \Xi_1^2 \Xi_2 } \left(2 \Xi_1 + 3 \Xi_2 \right)  $ &  $ \displaystyle -\frac{16 a_1 M \pi ^2 }{9} $ \\ 
  & $Q^{(2)}_{5}$ & $\ell^{a b c_1 c_2} x^{(1)}_{c_1} y^{(1)}_{c_2}$ & $\displaystyle  \frac{16 a_2 M \pi ^2 }{45 \Xi_1 \Xi_2^2 } \left(2 \Xi_2 + 3 \Xi_1 \right) $ & $\displaystyle  \frac{16 a_2 M \pi ^2 }{9} $ \\[0.5ex]
 \hline
 \multirow{2}{*}{6} & $Q^{(1)}_{6}$ & $\ell^{a b c_1 c_2 c_3} x^{(1)}_{c_1} x^{(2)}_{c_2} y^{(2)}_{c_3}$ & $ \displaystyle -\frac{5 a_1 M \pi ^3 }{48 \Xi_1 \Xi_2^2} \left(2 \Xi_1 + 3 \Xi_2 + \Xi_1 \Xi_2 \right) $ & $\displaystyle  -\frac{5 a_1 M \pi ^3 }{8} $ \\ 
  & $Q^{(2)}_{6}$ & $\ell^{a b c_1 c_2 c_3} x^{(1)}_{c_1} x^{(2)}_{c_2} y^{(1)}_{c_3}$ & $\displaystyle  \frac{5 a_2 M \pi ^3}{48 \Xi_1^2 \Xi_2} \left(2 \Xi_2 + 3 \Xi_1 + \Xi_1 \Xi_2 \right) $ & $\displaystyle  \frac{5 a_2 M \pi ^3}{8} $ \\ [0.5ex]
 \hline
  \multirow{3}{*}{7} & $Q^{(1)}_{7}$ & $\ell^{a b c_1 c_2 c_3 c_4} x^{(1)}_{c_1} x^{(2)}_{c_2} y^{(2)}_{c_3} y^{(3)}_{c_4}$ & $ \displaystyle \frac{16 a_1 M \pi ^3 }{175 \Xi_1^2 \Xi_2 \Xi_3}(2 \Xi_1 (\Xi_2 + \Xi_3)+3 \Xi_2 \Xi_3)$ & $\displaystyle  \frac{16 a_1 M \pi ^3 }{25}$ \\
  & $Q^{(2)}_{7}$ & $\ell^{a b c_1 c_2 c_3 c_4} x^{(1)}_{c_1} x^{(2)}_{c_2} y^{(1)}_{c_3} y^{(3)}_{c_4}$ & $ \displaystyle -\frac{16 a_2 M \pi ^3 }{175 \Xi_1 \Xi_2^2 \Xi_3}(2 \Xi_2 (\Xi_3 + \Xi_1)+3 \Xi_3 \Xi_1)$ & $\displaystyle - \frac{16 a_2 M \pi ^3 }{25}$ \\ 
   & $Q^{(3)}_{7}$ & $\ell^{a b c_1 c_2 c_3 c_4} x^{(1)}_{c_1} x^{(2)}_{c_2} y^{(1)}_{c_3} y^{(2)}_{c_4}$ & $ \displaystyle  \frac{16 a_3 M \pi ^3}{175 \Xi_1 \Xi_2 \Xi_3^2} (2 \Xi_3 (\Xi_1 + \Xi_2)+3 \Xi_1 \Xi_2)$ & $ \displaystyle  \frac{16 a_3 M \pi ^3}{25}$ \\ [0.5ex]
 \hline
  \multirow{3}{*}{8} & $Q^{(1)}_{8}$ & $\ell^{a b c_1 c_2 c_3 c_4 c_5} x^{(1)}_{c_1} x^{(2)}_{c_2} x^{(3)}_{c_3} y^{(2)}_{c_4} y^{(3)}_{c_5}$ & $ \displaystyle  \frac{7 a_1 M \pi ^4 }{288 \Xi_1^2 \Xi_2 \Xi_3}(2 \Xi_1 (\Xi_2 + \Xi_3)+3 \Xi_2 \Xi_3 + \Xi_1 \Xi_2 \Xi_3 )$ & $ \displaystyle  \frac{7 a_1 M \pi ^4}{36}$ \\
  & $Q^{(2)}_{8}$ & $\ell^{a b c_1 c_2 c_3 c_4 c_5} x^{(1)}_{c_1} x^{(2)}_{c_2} x^{(3)}_{c_3} y^{(1)}_{c_4} y^{(3)}_{c_5}$ & $ \displaystyle - \frac{7 a_2 M \pi ^4 }{288 \Xi_1 \Xi_2^2 \Xi_3}(2 \Xi_2 (\Xi_1 + \Xi_3)+3 \Xi_1 \Xi_3 + \Xi_1 \Xi_2 \Xi_3 )$  & $ \displaystyle -\frac{7 a_2 M \pi ^4}{36}$ \\ 
   & $Q^{(3)}_{8}$ & $\ell^{a b c_1 c_2 c_3 c_4 c_5} x^{(1)}_{c_1} x^{(2)}_{c_2} x^{(3)}_{c_3} y^{(1)}_{c_4} y^{(2)}_{c_5}$ & $ \displaystyle \frac{7 a_3 M \pi ^4 }{288 \Xi_1 \Xi_2 \Xi_3^2}(2 \Xi_3 (\Xi_1 + \Xi_2)+3 \Xi_1 \Xi_2 + \Xi_1 \Xi_2 \Xi_3 )$ & $ \displaystyle  \frac{7 a_3 M \pi ^4 }{36}$ \\ [0.5ex]
\end{tabular} 
\renewcommand{\arraystretch}{1}
\end{table}
\end{landscape}
}

We find that the generic expression for ${\cal Q}^{(j)}_{D}$ can be written as
\beq
\label{QGLPP}
\mbox{GLLP:} \quad {\cal Q}^{(j)}_{D} = \frac{(D-1) \, \Omega(D-1)}{2D (D-2)} \, 
\frac{p^{(j)}_{D} (\lambda)}{\Xi_j \, \left( \prod_{k=1}^{N} \Xi_k \right)} \, a_j M \,, \qquad
\Omega(D-1) := \frac{2\pi^{D/2}}{\Gamma(D/2)} \,.
\eeq
$\Omega(D-1)$ is the surface area of the unit sphere $S^{D-1}$.
Here $p^{(j)}_{D}(\lambda)$ is a polynomial that contains $\Xi$ terms, which can
be written iteratively starting from the first nontrivial one $p^{(1)}_{4}$. For $D \geq 5$,
$p^{(1)}_{D} (\lambda)$ explicitly reads
\beq
p^{(1)}_{D} (\lambda) = p_{2N}^{(1)}(\lambda) \, \Xi_{N} + 
2 \prod_{k=1}^{N-1} \, \Xi_{k} - (1-\epsilon) \prod_{k=1}^{N} \Xi_{k} \, 
\quad \mbox{with} \quad 
p^{(1)}_{4} := 3 + \Xi_{1} \,.
\eeq
The remaining $p^{(j)}_{D}(\lambda)$, for $j = 2, \dots, N$, follows by cycling the $N$ 
indices that $j$ runs over. For the MP metrics for which \( \lambda = 0 \),
\( \Xi_{j} = 1 \) and \( p^{(j)}_{D}= D \) for all $j$, so that \re{QGLPP} simplifies to
\beq
\label{QMP}
\mbox{MP:} \quad {\cal Q}^{(j)}_{D} = \frac{(D-1) \, \Omega(D-1)}{2D (D-2)} \, a_j M \,.
\eeq
The charges ${\cal Q}^{(j)}_{D}$ \re{QGLPP} are proportional to $a_j M$ and are clearly
related to the angular momenta of the black holes. The correct angular momenta that
fulfills the first law of black hole thermodynamics have been calculated through the Komar 
integral and reads\footnote{See Section 4.1 of \cite{Gibbons:2004ai} for details.}
\beq
J_j = \frac{\Omega(D-2)}{4 \pi \, \Xi_j \, \left( \prod_{k=1}^{N} \Xi_k \right)} M a_j 
\label{JGary}
\eeq
in our conventions. The main difference stems from the polynomial $p^{(j)}_{D}(\lambda)$ 
that our ${\cal Q}^{(j)}_{D}$ \re{QGLPP} has. However, recall that our main objective is to
come up with some conserved charge out of the KT-current. In that sense, we never expected
${\cal Q}^{(j)}_{D}$ \re{QGLPP} to have a clear-cut or a definite physical meaning in the first
place.

\section{Discussion}
\label{disc}
In this paper we first reviewed and amended the construction of asymptotic charges 
starting from the KT-current for KYTs. We then applied these ideas to the GLPP and MP 
black holes to arrive at the charge formulae \re{QGLPP} and \re{QMP}, respectively.

The construction we have presented heavily relies on a set of properly chosen
vectors $x^{(i)}$, which in itself implicitly depends on the foliation of the background.
This naturally brings in the question of whether the charge ${\cal Q}$ \re{theQ} is 
background gauge invariant, i.e. how does the charge ${\cal Q}$ \re{theQ} change as
the deviation $h_{ab}$ \re{metspl} transforms as
\beq
 \delta_{\bar{\zeta}} h_{ab} = \bar{\nabla}_{a} \bar{\zeta}_{b} + \bar{\nabla}_{b} \bar{\zeta}_{a} \label{htra}
\eeq
under an infinitesimal diffeomorphism generated by a vector $\bar{\zeta}^{a}$? The answer
to this question, foremost, depends on how the linearized current $(j)_{L}$ \re{KTComL} 
itself transforms under \re{htra}. Using \re{RiemL}, we find for an arbitrary background that 
\beq
 \delta_{\bar{\zeta}} (R_{ab}{}^{cd})_{L} = 
 2 \bar{R}_{ab}{}^{e[c} \bar{\nabla}_{e} \bar{\zeta}^{d]} 
 - 2 \bar{R}^{cd}{}_{e[a} \bar{\nabla}_{b]} \bar{\zeta}^{e} \,.
  \label{Riemtra}
\eeq
However, for maximally symmetric backgrounds we have been working with 
the right hand side vanishes so that \( \delta_{\bar{\zeta}} (R_{ab}{}^{cd})_{L} = 0 \), 
which means, via \re{KTComL}, that the linearized current $(j)_{L}$ and hence its 
potential $\bar{\ell}$, via \re{Rdiff}, are both background gauge invariant\footnote{The 
potential is left invariant up to an exact term.}. Since the infinitesimal 
diffeomorphisms in question do not change the foliation of the background, the 
vectors $x^{(i)}$ are also left invariant. Thus we conclude that the charge 
${\cal Q}$ \re{theQ} is background gauge invariant for maximally symmetric backgrounds.

Our main aim has been to associate a conserved charge to the enigmatic
KT-current. In that sense, we feel our work gets to first base.

\bigskip

\noindent{\bf Acknowledgments}\\
The research of U.L. is supported in part by the 2236 Co-Funded 
Scheme2 (CoCirculation2) of T\"UB{\.I}TAK (Project No:120C067)\footnote{\tiny However 
the entire responsibility for the publication is ours. The financial support received from 
T\"UB{\.I}TAK does not mean that the content of the publication is approved in a scientific 
sense by T\"UB{\.I}TAK.}, and in part by The Leverhulme Trust.
\bigskip

\end{document}